Experimental confirmation of the negentropic character of the diffraction polarization of diffuse radiation

V. V. Savukov

In the course of analyzing the axiomatic principles on which statistical physics is based, the assumption of the limited correctness of the postulate that all allowable microstates of a closed system are equally probable was checked. This article reports the results of a study of a quasi-equivalent system within which isotropic radiation interacts with a phase diffraction grating. A simulated computer model of such interaction showed that anisotropic polarization must arise in the diffracted radiation, which reduces the Boltzmann entropy of the entire system and allows an external observer to obtain information on the grating's surface topology. This prediction was confirmed when it was experimentally checked directly on actual apparatus.

## Table of contents

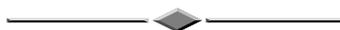

# Introduction

The diffraction mechanism by which a gas of quantum particles undergoes scattering makes it theoretically allowable for nonergodic systems to exist whose behavior lies outside the "zone of accountability" of statistical physics. There is a basis for assuming, for example, that a diffuse monochromatic photon gas that undergoes scattering at an ideally conducting reflective phase diffraction grating can alter its initial isotropic macrostate to an anisotropic one. If the photon gas and the diffraction grating are components of a closed physical system, the Boltzmann entropy of the given system must spontaneously decrease during such a process [1].

This article presents the results of direct experimental checking of the theoretical prediction made above, carried out on an actual physical apparatus.

# Formulation of the problem

Let there be a closed system in the form of a cavity that has a small aperture and is filled with diffuse radiation. A diffraction grating that interacts with the given radiation is also placed in the cavity. The result of the planned experiment is to theoretically predict that $S$-polarization components can be detected in the radiation scattered by the grating whose radiance characteristics exceed those of the components in the initial diffuse radiation [1]. The scattering index in this case must simultaneously contain similarly located (including the range of reflection angles) $P$-polarization components that have substantially lower radiance.

The chief complication of the experimental checking indicated above is the need to differentiate the effects caused by the simultaneous manifestation of Rayleigh–Wood resonance and threshold anomalies.

Rayleigh–Wood resonance anomalies characterize the absorption zones of the radiation incident on the grating. The threshold anomalies correspond to regions in which the efficiency of one radiation component or another (for instance, the polarization components) is redistributed between various diffraction orders of single-photon scattering. If the scale of the resonance anomalies substantially predominates over the threshold anomalies, it will be impossible to reliably identify the manifestation of the desired effect [2].

A single "window of possibility" for which the predicted effect must be fairly intense has been detected in the region of visible radiation. This is a version of the interaction of diffuse monochromatic radiation (wavelength at least 700 nm) with a gold diffraction grating having a sinusoidal profile with the following parameters:

– the ratio of the step of the grating to the radiation wavelength is $d/\lambda = 0.82$,

– the ratio of the grating's total relief depth to its step is $h/d = 0.38$.

The diffuse-radiation surface absorbance of such a grating does not exceed 0.069 for $\lambda = 700$ nm. This is acceptable from the viewpoint of performing the proposed task, since the appearance of Rayleigh–Wood resonance anomalies has a rather moderate character for such an absorbance.

# Description of the experimental apparatus

Figure 1 shows the layout of the apparatus. The principle that it uses to obtain a three-dimensional image is known as the fisheye effect, in which the visual parameters of the object are recorded at the same time for all the possible combinations of the observation angles. This principle is based on photographing the surface reflection from the diffraction grating in a spherical mirror located close to this grating. The picture is taken through a small aperture in the body of the grating itself. Noise can be observed in the middle and to the right and left on the frames of the photographs in the form of three secondary reflections of the spherical mirror: the central one, formed by the zeroth order of the diffraction grating, and the two side ones, formed by the ±1 orders [3].

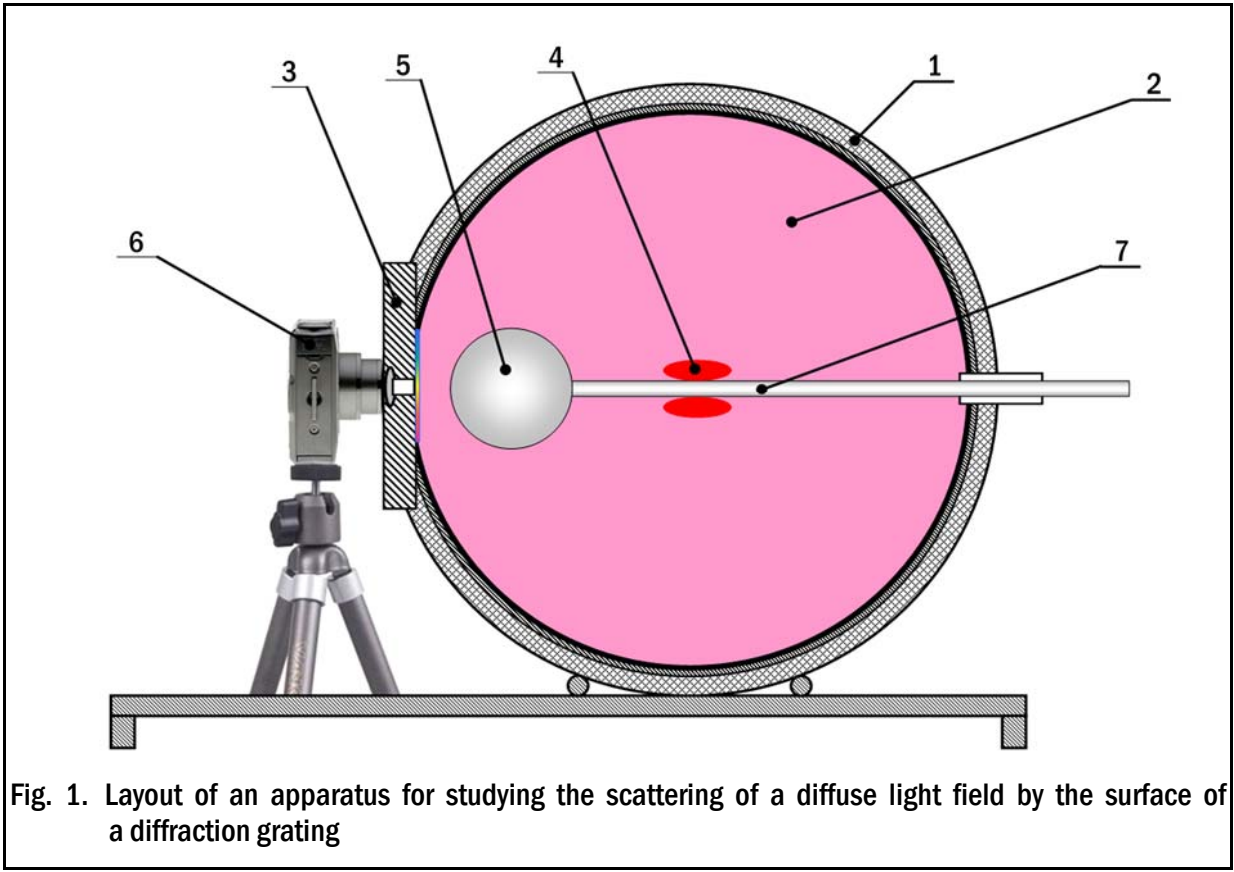

Fig. 1. Layout of an apparatus for studying the scattering of a diffuse light field by the surface of a diffraction grating

## The apparatus (Fig. 1) includes the following elements:

1. A hollow photometric sphere, which models a quasi-closed system.

2. The inner surface of sphere 1, coated with white matte paint having a spectral reflectance of about 0.97.

3. A flat reference shim on whose inner surface (turned toward the cavity of sphere 1) is attached a disk-shaped flexible replica diffraction grating with a gold coating; shim 3 and the replica grating have a circular aperture in front of the objective of camera 6;.

4. LEDs with quasi-monochromatic radiation ($\lambda \approx 700\pm50$ nm).

5. A spherical mirror that serves for three-dimensional scanning of the diffraction grating attached to reference shim 3.

6. A camera intended for obtaining three-dimensional pictures of the grating surface; an analyzer is mounted in front of the camera lens when the polarization components are being recorded.

7. A rod that serves to fix LEDs 4 and spherical mirror 5.

## Experimental organization and results

Before the actual physical experiment is carried out, simulation modeling is used to predict the expected effect. We considered an isolated optical system that contains a monochromatic ($\lambda \approx 700$ nm) stochasticized diffuse photon gas that interacts either with an ideally conductive diffraction grating or with a gold reflective phase-type diffraction grating that has one-dimensional (linear) sinusoidal surface microrelief (the number of photons in the statistical experiment is $2^{20} - 1 = 1,048,575$).

The macroscopic parameters of the desired light field formed as a result of the given interaction were computed on the basis of the corresponding theoretical model. Software was specially created earlier and was verified for simulation modeling of the processes by which a diffuse photon gas is scattered on various forms of reflective optical elements (see [3-5] for details).

The predicted character of the scattering of various components of the diffuse light field on a gold diffraction grating must then be compared with the corresponding results recorded on the physical apparatus. However, a simple visual comparison of graphical images of the index obtained in the course of simulation and actual experiments cannot give an unambiguous answer concerning the presence or absence of the desired effect. In order to eliminate the probability of interpreting these data subjectively, some quantitative criterion should be used that makes it possible to estimate whether the experimental results are valid.

Quantitative monitoring of the radiance of the *S*-polarization component of the scattered radiation was chosen as the indicated criterion. It is predicted that, even for an actual gold grating with finite conductance, the indicated radiance for individual solid angles must exceed half the ideal value (with no losses to absorption) of the Lambertian level. Since one half of an arbitrarily chosen flux of stochasticized diffuse photon gas has to be in each of the *S* and *P* components, an excess of this "half" level in either of the polarization components can no longer be compensated by the other component. Consequently, the total diffracted radiation flux unavoidably acquires an anisotropic polarization structure. This conclusion remains valid even when its intrinsic thermal radiation, determined by Kirchhoff's law from the condition of thermodynamic equilibrium, is included in the radiation flux that leaves the grating [1].

Figure 2 illustrates the methodology for determining the numerical criterion that corresponds to the relative radiance of the reflection of a diffuse light field from an ideally conducting mirror (i.e., with no losses to absorption). The calibration is carried out on an actual gold mirror for radiation with $\lambda \approx 700$ nm.

Figure 2(a) shows a photograph of the three-dimensional reflection index of a monochromatic diffuse light field from the gold mirror. The index is constructed in a polar coordinate system in such a way that its center corresponds to the zeroth reflection angle with external scanning of the surface. The reflection angle is proportional to the polar radius and approaches 90° at the periphery of the circular diagram. This site is visible on the photograph as the annular interface of the mirror with the inner surface of the photometric sphere (see the diagram of the apparatus in Fig. 1). The azimuthal angle of observation of the grating surface is determined by the polar angle of the diagram. Secondary reflection of the spherical mirror is seen in the central part of the photograph (see Fig. 1, position 5), formed by the zeroth order of scattering, which is single in this case. High-contrast reflection of the dark circular aperture in the reference shim is also clearly seen (see Fig. 1, position 3), and the picture was taken through this aperture.

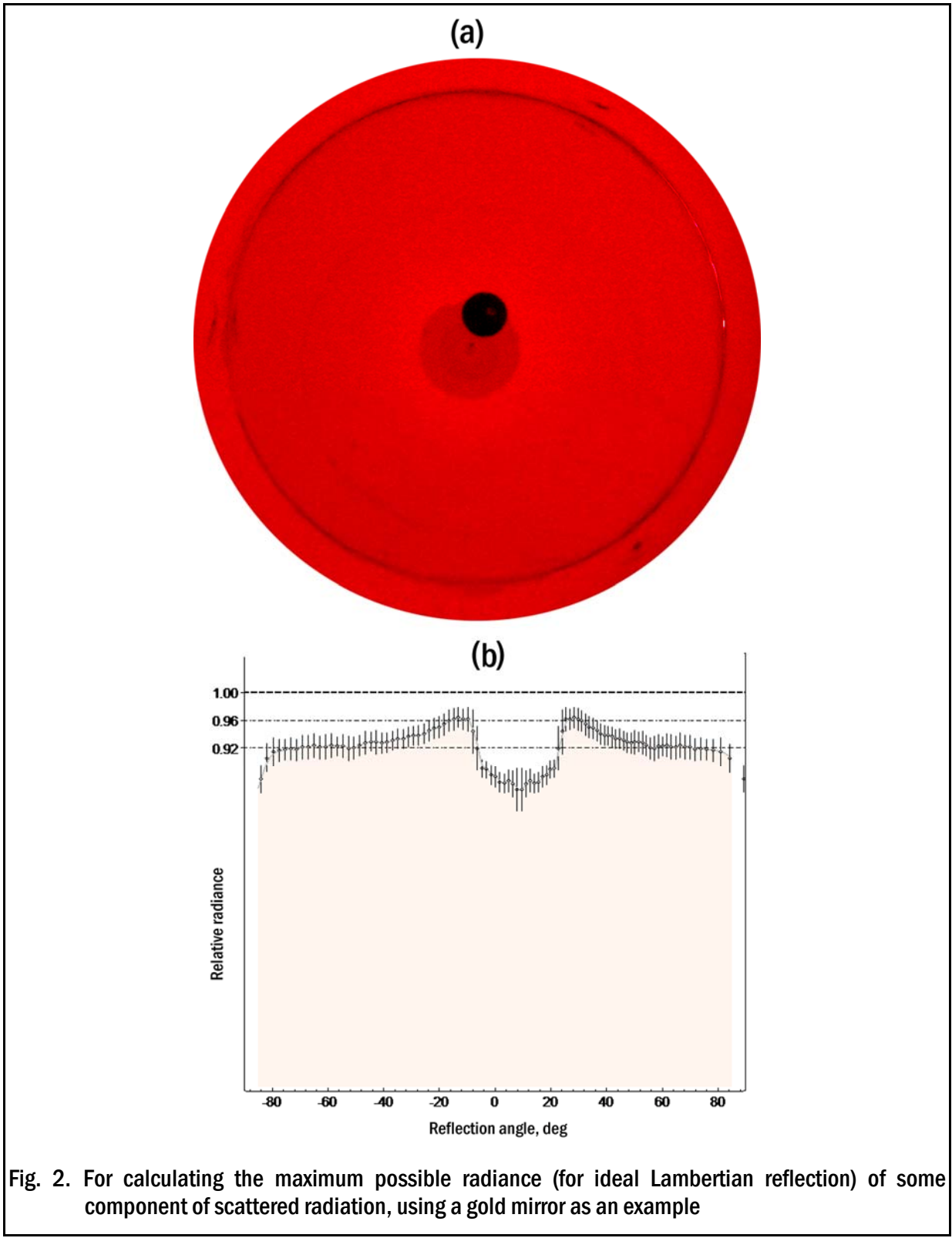

Fig. 2. For calculating the maximum possible radiance (for ideal Lambertian reflection) of some component of scattered radiation, using a gold mirror as an example

Under the photograph of the index [Fig. 2(a)] is a graph [Fig. 2(b)] that shows how the relative radiance of the radiation (vertical axis) depends on the reflection angle, which can vary from −90° to 90° in the plane of the horizontal cross section of the scattering index (horizontal axis). The reference radiance labels shown by thin dashed lines on the graph cor-respond to the calculated reflectance values of 0.96 for the case of normal reflection ($\theta \approx 0°$) and 0.92 for a glancing angle ($\theta \to \pm 90°$). The single level of the relative radiance that corresponds to the ideal Lambertian reflection with no losses to absorption is shown by a thicker dashed line. The level for this ideal case is computed by means of linear extrapolation of the data, using the already established reference levels. Note that this approximation gives good results only in the nearest neighborhood of the existing reference points. Such a limitation makes it meaningless to attempt to completely digitize the entire vertical axis (the relative radiance), especially since there is no real need to do it.

The relative radiance levels on the graphs of the horizontal cross sections of the scattering index of the $S$ and $P$ polarization components are computed similarly. The level that corresponds to ideal Lambertian reflection with no losses to absorption equals one half for each component in this case. For diffraction gratings, special attention should be paid to the selection of the reference zones on the scattering index. These zones were specified on the diagonal sections of the diagrams, where Rayleigh–Wood resonance anomalies and noise in the form of secondary reflections are minimal.

Figure 3 shows images of the total radiance indices [Fig. 3(a)] and the polarization components of the scattered light field [the $S$ component in Fig. 3(b) and the $P$ component in Fig. 3(c)], obtained for an ideally conductive linear phase grating (step $d = 588$ nm, total depth of the sinusoidal microrelief profile $h = 223$ nm). The macroscopic radiance gradients of the $S$ and $P$ polarization components of the scattered radiation separated by software [Figs. 3(b) and 3(c)] are visible on the diagrams. The expected radiance difference of the actual physical light field is predicted in this case to be at about 5%–6% of the value corresponding to Lambertian reflection. Such a pronounced effect is caused by the presence of Rayleigh–Wood threshold anomalies in the complete absence of resonance anomalies (an ideally conducting grating). The radiances of the $S$ and $P$ components complement each other so that their sum is ordinary unpolarized diffuse radiation [see Fig. 3(a)].

The next row of Fig. 3 shows the total radiance index [Fig. 3(d)] and the polarization components [the $S$ component in Fig. 3(e) and the $P$ component in Fig. 3(f)], obtained by modeling the diffuse radiation scattering at a gold grating with the same microrelief parameters as for the case of ideal conductivity described earlier. Here the predicted effect shows up in the form of a horizontal bright band that passes through the central part of the diagram of the $S$ component and its corresponding dark band on the diagram of the $P$ component.

Figures 3(g)–3(i) show pictures of the scattering diagrams (which have been gamma-corrected), obtained in a physical experiment on a gold diffraction grating with the same microrelief parameters as in the gratings used earlier in the computer models. The graph that appears under each diagram of this series shows how the relative radiance of the observed component of the radiation (the vertical axis) depends on the reflection angle, which can vary from $-90°$ to $90°$ in the plane of the horizontal cross section of the scattering index (the horizontal axis). The dashed lines in the graphs show the radiance levels of the various components that correspond to ideal Lambertian scattering with no losses to absorption: $R = 1$ [Fig. 3(j)] and $R = ½$ [Figs. 3(k) and 3(l)].

The fact that this ideal radiance level predominates in the experimentally recorded $S$ component [see Figs. 3(h) and 3(k)] and lies in the region of those angles where the predicted effect was expected to be most intense [see Figs. 3(b) and 3(e)] makes it possible to say that reality matches prediction fairly well.

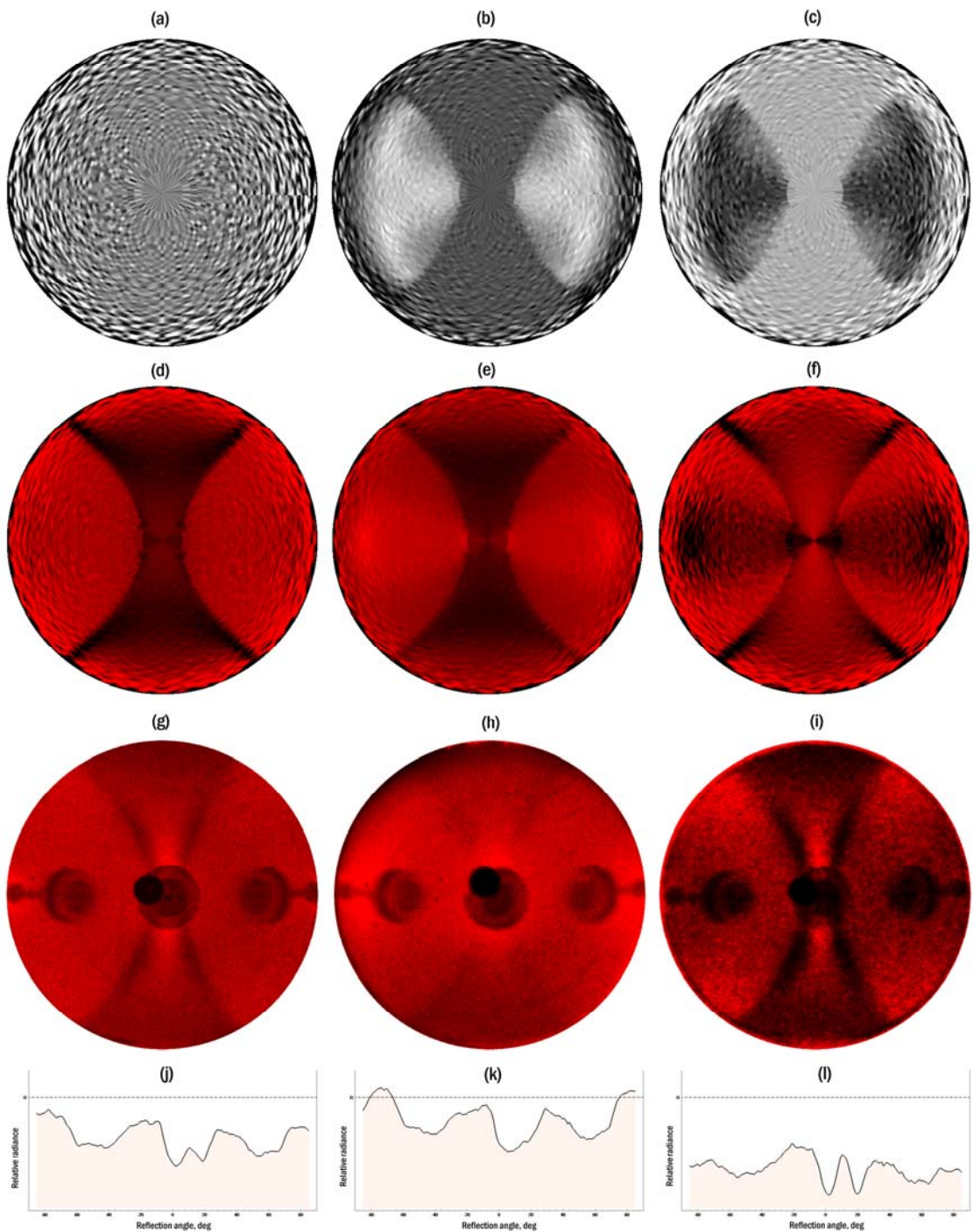

Fig. 3. - Angular diagrams of the total radiance and of the polarization components of a diffuse light field scattered by the surfaces of the following diffraction gratings: (1) with ideal conduction (computer model, maximum deviation of the *S* and *P* components from the ideal Lambertian level 5%–6%); (2) with gold coating (computer model, excess of the ideal Lambertian level of the *S* component as much as 1.5%–2%); (3) with gold coating (actual experiment, excess of the ideal Lambertian level of the *S* component 1%–1.5% in the angular directions that correspond to the calculated maximum in the computer model).

It is useful to inquire how accurately the technical conditions must be obeyed in order to successfully perform the described experiment. Along with the step size of the diffraction grating, it is important to maintain the required depth, as well as its sinusoidal microrelief profile. Example: Fig. 4(a) shows a sinusoidal grating with the necessary step and depth. The diagram for the *S* component of the scattering in Fig. 4(b) shows that the desired effect is present. The deviation from the sinusoidal profile in Fig. 4(c) (with the same step and depth) causes the effect to disappear [Fig. 4(d)].

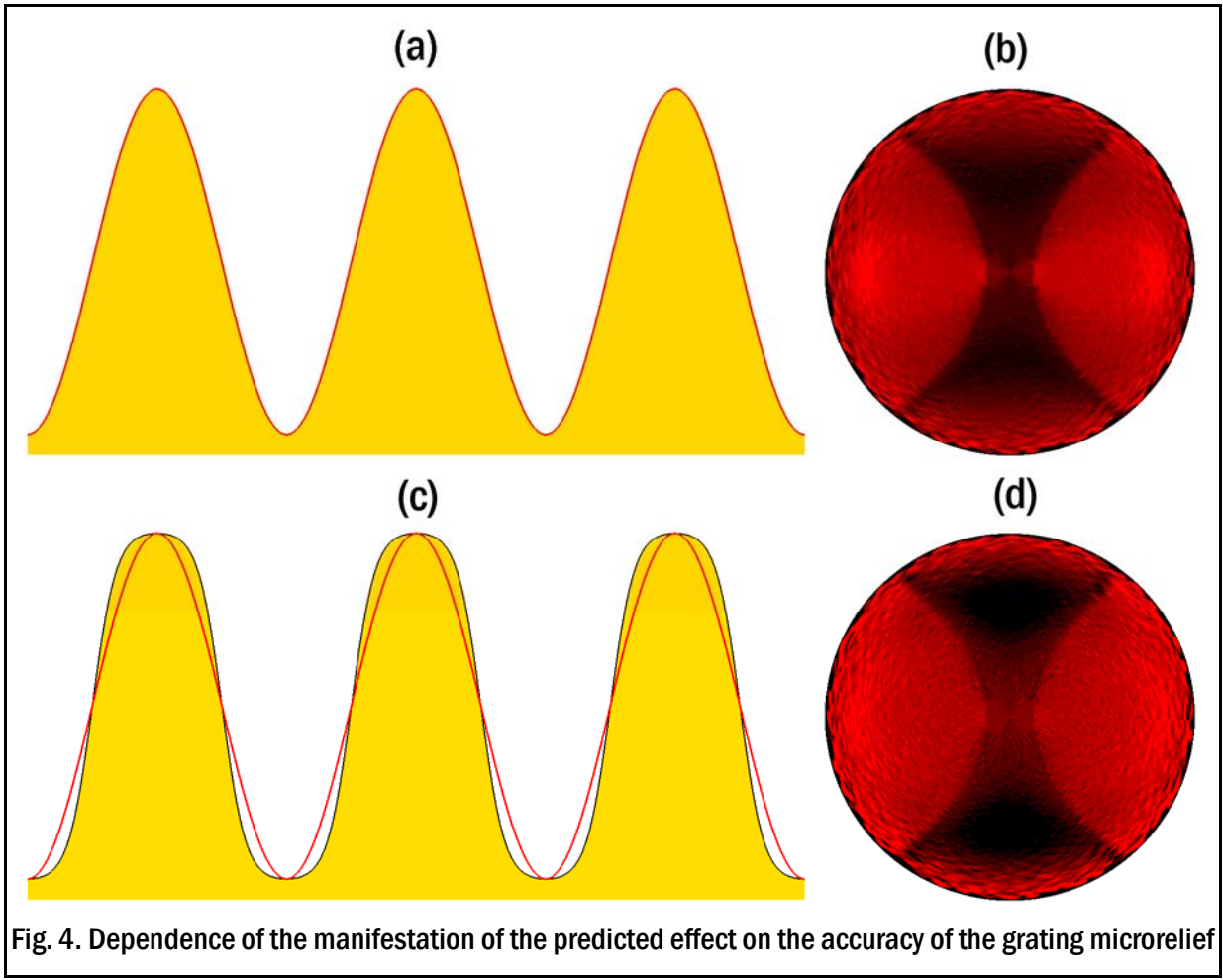

**Fig. 4.** Dependence of the manifestation of the predicted effect on the accuracy of the grating microrelief

## Conclusions and prospects for further studies

It has been experimentally confirmed that the monochromatic components of a diffuse photon gas can acquire anisotropic polarization in the process of diffraction scattering implemented in a closed physical system.

Thus, for example, blackbody radiation can exhibit inhomogeneous polarization structure when it interacts with a diffraction grating from which this radiation was initially in a state of thermodynamic equilibrium. This provides a basis for revising the concept of the most probable macroscopic state of a closed system, which is based on the postulate that its microstates are equally probable. In particular, the actual macrostate can be adequately described by using entropy defined via Shannon's formula [6]. This makes it possible to formulate the concept of entropy as a measure of the probability of some macrostate of a closed system without having recourse to the basic postulate of statistical physics indicated above, used with the definition of Boltzmann entropy.

These results undoubtedly need to be carefully checked. If their reliability is confirmed, they will become important in practice. For example, technical decisions are obvious if they are associated with the task of passive direction-finding of an object "marked" by a diffraction grating and in thermodynamic equilibrium with the surroundings (for concealed security systems, etc.).

**Funding.** Ministry of Education and Science of the Russian Federation (grant: 9.1354.2014/K).

**Acknowledgment.** The author is deeply grateful to Igor Golubenko, who actively participated in creating the software used in this project.

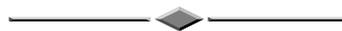